# Unveiling Magnon-Magnon Coupling and Its Dynamic Control in Nanomagnets

**Siddhesh Sharad Kashid, Sachin Verma, Abhishek Maurya, Manjushree Maity, Kuldeep Kumar Shrivastava, Rajeev Singh and Biswanath Bhoi***

*Nano-Magnetism and Quantum Technology Laboratory, Department of Physics, Indian Institute of Technology (Banaras Hindu University) Varanasi, Varanasi - 221005, India*

Hybrid magnonics, exploring the coupling between magnons and quantum systems, is an exciting field for developing next-generation information technologies. Achieving a strong and tunable magnon-magnon coupling (MMC) in confined nanomagnets is crucial for the on-chip integration of these hybrid systems and advancing the field. In this work, we numerically investigate the interactions between different magnon modes excited within an elliptical magnonic nano-disc (EMND), demonstrating an anti-crossing effect in the dispersion spectra. A comprehensive theoretical framework was presented that explains this anti-crossing phenomenon as a result of MMC and provide estimates for the strength of the coupling ($g$). Furthermore, we show that this intermodal coupling can be tuned from a strong coupling regime ($g$ = 300 MHz) to a weak coupling regime by varying the direction of the external magnetic field and the intrinsic properties of the EMND. Our combined numerical and theoretical findings offer new insights into MMC, significantly advancing the field of quantum magnonics and magnon-based quantum information technology.

# 1. INTRODUCTION

Magnons, the quanta of spin waves (SW) representing collective excitations of spins in magnetic materials, has been widely investigated for energy-efficient and high-speed technologies due to its charge-less diffusion and long coherent distance and time [1-2]. Recently 'Hybrid Magnonics' an emerging research field based on magnons has garnered widespread attention due to its great potential for quantum information processing devices [3-5]. It explores the coherent transfer and storage of information between magnons and various elemental excitations (such as photons, phonons, etc) enabled by their strong hybridization with magnons [6]. Hybrid systems relying on magnon-photon or magnon-phonon coupling, typically using microwave cavities and magnetic materials in the millimeter range, face challenges related to device miniaturization and integration with complementary metal-–oxide–semiconductor (CMOS) technology [3-7]. Therefore, the pursuit of hybrid quantum devices that can be downscaled to the nanoscale while preserving strong coupling has become a highly active area of research.

Over the year some studies have identified that magnons can interact with other magnons in multilayer or hetero-structure system precisely has the advantages of scaling down to the nanoscale while retaining strong coupling. Klingler et al and Chen et al independently discovered the robust interlayer magnon-magnon coupling (MMC) in spatially separated metal-insulator hybrid multilayers, including YIG/FeNi heterojunctions [8] and synthetic antiferromagnetic CoFeB/Ru/CoFeB structures [9], while Li et al [10] provided additional clarification of the physical mechanism. However, coupling between the localized SW modes within the same nanomagnetic system can have the advantage of superior coupling strength compared to other hybrid quantum systems, as it mitigates the reduction in coupling strength typically caused by insufficient spatial mode overlap. Understanding these interactions is crucial for developing spintronic devices, where controlled spin wave propagation and interference can be used for information processing at the nanoscale. Moreover, the demonstration of MMC between two modes within a common host medium can enable device miniaturization and integration with CMOS technology for quantum information processing.

In light of these prevailing research gaps, Dai et al [11] and Ye et al [12] showed the interaction between different magnon modes excited in a single hexagonal nanomagnet by micromagnetic simulation while the group led by Barman et al [13] experimentally demonstrated MMC in cross-shaped nanomagnets [14], magnetic nanocross arrays [15] and nanoring [16]. As the foundation for quantum information processing is the control of quantum excitations and their coherent interactions, further emphasis must be given to the tunability and controllability of magnon-magnon interactions in nanoscale systems. Few studies have explored the manipulation of SW dynamics in nanomagnets by tuning intrinsic parameters like dimensions (shape, size, symmetry) and material choice, as well as extrinsic factors like the strength and orientation of the bias magnetic field. While in-plane bias field effects have been examined [11, 12, 15], the influence of out-of-plane fields, along with intrinsic magnetic properties such as anisotropy, damping, and magnetization, remains unexplored. Given the relatively limited exploration of MMC in single nanomagnets, there is a need to expand research on MMC using various geometries and systems. Our study aims to achieve and elucidate the MMC phenomenon in some new geometry, shedding light on the underlying mechanisms of hybrid magnon interactions and their tunability in a much more detailed fashion.

In this study, we performed a numerical simulation to investigate the interactions among different magnon modes confined within an elliptical magnonic nano-disc (EMND). We identified a distinct signature of coupling, characterized by anticrossing in the frequency-field (f-H) dispersion relations, which is attributed to the interaction between edge and center magnon modes. Significantly, the occurrence of anticrossing was found to be highly sensitive to the orientation of the magnetic field relative to the EMND geometry. By varying the angle between the external magnetic field and the z-axis of the nanomagnet, we observed that coupling is most pronounced at angles between 30º and 60º. This observation is supported by an analysis of the simulated magnetic field distributions and phase maps of various spin-wave modes within the system, both near and far from the anticrossing point. Our results suggest that this system could serve as a novel magnonic platform for enhancing the exchange of quantum information between strongly coupled magnons, thereby advancing the development of magnon-based quantum information processing technologies.

## 2. DESIGN AND SIMULATION SETUP

To compute the spatial magnetization distributions as well as the magnetization dynamics in ferromagnetic nano-disc. we employed the Mumax3 code that solves Landau-Lifshitz-Gilbert equation as given by [17]

$$\frac{\partial \vec{M}}{\partial t} = -\gamma_G (\vec{M} \times \vec{H}_{eff}) + \frac{\alpha_G}{M_0}\left(\vec{M} \times \frac{\partial \vec{M}}{\partial t}\right)$$

where $\gamma_G$ is gyromagnetic ratio, $\alpha_G$ is the Gilbert damping constant and $\vec{H}_{eff}$ is the effective field obtained by the functional derivative of total E given as $\vec{H}_{eff} = -\left(\frac{1}{\mu_0 M_0}\right)\left[\frac{\delta \tilde{E}(\vec{m})}{\delta \vec{m}}\right]$. Here $\tilde{E}$ is the total sum of exchange energy density, dipolar energy density, anisotropy energy density (if any) and Zeeman energy density. $\mu_0$ is the magnetic permeability of free space. The reduced magnetization, denoted as $\vec{m}$ is calculated by dividing the magnetization vector $\vec{M}$ by the saturation magnetization $M_0$.

The geometrical parameters illustrated in Fig. 1 depict that the EMND are positioned on the x-y plane, with the thickness oriented along the z-axis. For a model system of FeNi EMND, we used the following material parameters [15, 16]: $M_s$ =1.2×10$^6$A/m, $\alpha_G$ = 1 ×10$^{-3}$, $\gamma_G$ = 17.95 MHz Oe$^{-1}$, exchange stiffness constant $(A_{ex})$ = 1.1×10$^{-11}$J/m and uniaxial anisotropy constant $K_u$= 5 kJ/m$^3$ along the width of the magnonic cavities. Two step simulation was performed for different values of external bias field $(H_{ext})$ to obtain the Magnon Spectra. First, a static simulation was carried out to get the ground state of the magnetization determined by minimizing the total energy of the simulated magnetic volume. Second, a dynamic simulation starting from the ground magnetization state was done with a radio frequency perturbation field $(h_{rf})$ applied perpendicularly to the $H_{ext}$, as schematically shown in Fig. 1. The $h_{rf}$ was adapted in the form of a "Sinc" function [17, 18], $h_{rf}(t) = h_0 \frac{Sin(2\pi ft)}{(2\pi ft)}$, where the amplitude $h_0$ = 5 mT and the cut-off frequency $f$ = 50 GHz. The magnons with frequencies ranging from 0 to 50 GHz can be effectively excited. During the dynamic simulation of 10 ns, the spatially averaged magnetization $m(t)$ was saved with a time interval of 10 ps. Then, the magnon spectra at the specific field can be obtained by performing Fourier transform of the recorded $m(t)$.

## 4. RESULTS AND DISCUSSION

### 4.1. Observation of Anti-crossing

Figure 2 (a) illustrates the color-mapped frequency dependence of magnon modes as a function of the external bias field H applied at various angles θ = 0, 15, 30, 45, 60, 75 and 90°. This figure highlights the intricate behavior of magnon modes under varying magnetic field orientations, providing insights into the dynamic interactions within the system. Notably, a distinct anti-crossing gap emerges at angles θ = 15°, 30°, 45° and 60°, signifying the coexistence of two spin-wave modes within certain field strengths. This coexistence is a key indicator of magnon-magnon coupling, where the two modes interact and influence each other's behavior. The anti-crossing phenomenon is particularly pronounced at angles $\theta = 45°$ and $60°$, where the coupling is strongest, indicating a significant interaction between the SW modes. In contrast, this effect is less noticeable at $\theta = 0°$, suggesting weaker coupling, and is entirely absent at $\theta = 90°$, where the external field orientation does not favor such interactions. Outside these regions of coupling, the frequency curves exhibit remarkable similarity, with the trajectories of the magnon modes nearly overlapping, indicating that the magnon dynamics are largely independent of the external field angle in these areas. This observation underscores the specific conditions under which magnon-magnon coupling occurs. In light of these significant findings, our investigation seeks to further explore the underlying physics governing these interactions and to identify potential applications arising from these observations.

To gain a deeper understanding of the strong MMC phenomena, we carefully examine the representative magnon spectra at various field strengths for $\theta = 60°$, as depicted in Fig. 3(a). At a field strength of H = 180 mT, the lower frequency mode (M1) dominates in intensity over the higher frequency mode (M2), indicating a stronger excitation or greater population of the M1 mode at this lower field value. This observation suggests that, at this point, the system's magnetic configuration favors the M1 mode, possibly due to the alignment of magnetic moments or the influence of magnetic anisotropy. As the external magnetic field H is progressively increased from 180 to 320 mT, a notable shift in the spectral characteristics occurs: the intensity of the M1 mode diminishes, while the intensity of the M2 mode correspondingly increases. This inverse relationship between the intensities of M1 and M2 modes indicates a dynamic energy transfer or redistribution between the two magnon modes, likely driven by the evolving magnetic environment as the field strength increases. By the time the field reaches H=320 mT, the M2 mode overtakes the M1 mode in intensity, signifying a reversal in the dominance of the magnon modes. This transition is a clear indication of the field-dependent nature of MMC, where the external field can significantly alter the energy landscape, thereby modulating the relative populations of the magnon modes. To understand in more detail, we investigated the spatial profiles of dynamic magnetization near magnon-magnon interactions. The dynamic simulations generated output files containing information about the spatial and temporal distribution of magnetization. By applying a discrete fast Fourier transform (FFT) to the datasheet, we extracted and plotted the power values at the desired frequency to create the system's power profile. Figure 3(b) presents the amplitude distribution of magnetization components corresponding to two interacting magnon modes M1 and M2 at five distinct field values after the Fourier transform. At H = 320 mT, mode M1 with a frequency a $f_1$ = 12.7 GHz exhibits resonance mainly at the upper and lower edges, identifying it as the edge mode. Stronger resonance is indicated by colors closer to red. In contrast, mode M2, also at $f_2$ = 14.1 GHz, shows resonance primarily in the center and aligns parallel to the edges, characterizing it as the center mode. As we move toward regions of strong coupling, the color intensity for mode M1 increases both at the edges and the center, whereas the intensity for mode M2 decreases. At H = 193 mT, both modes M1 and M2 display nearly equal color intensities at both the edge and center, suggesting that resonance is distributed between these regions in the EMND. However, at H = 180 mT, mode M1 with $f_1$ = 10.0 GHz shows dominant resonance in the center, while mode M2 with $f_2$ = 10.8 GHz exhibits stronger resonance at the edges compared to the center. These observations confirm that the energy exchange between the edge and center modes within the EMND, Therefor the MMC in EMND arises from the interaction between the spin wave center mode and the edge mode, primarily driven by dipole–dipole interactions. This behavior is consistent with previous studies, such as those by Keatley et al. on single square nanomagnets [20], Lihui Bai et al. on FeNi microstrips [21], and Changting et al. on FeNi hexagons [11], which reported similar mode transitions and coupling dynamics. The agreement with these studies reinforces the generality of the MMC phenomena across various magnetic structures and materials.

Additionally, it is important to note that the critical transition point, or turning point, in the relative intensities of the M1 and M2 modes occurs at H = 193 mT, where both modes exhibit nearly equal intensities, indicating a balance between the two. This specific field value is defined as the coupling field $H_g$, where the MMC is most effective, resulting in almost equal energy exchange between the modes. At this coupling field, the frequency difference between the M1 and M2 modes reaches its minimum value, indicating a resonance that characterizes strong MMC and enhances coupling efficiency. More importantly, as the angle θ increases from 0, the anti-crossing or coupling center or the coupling field $H_g$ shifts progressively to lower magnetic field values (Fig. 2(b)), eventually becoming undetectable at θ = 90°. This suggests a possible correlation between the applied field angle and the observed magnon-magnon interactions.

We further explored the reasons behind the coupling occurring only within a specific range of angles and the observed decrease in the coupling field as the applied magnetic angle increases. This behavior indicates that the energy landscape of the SW modes becomes more favourable to interaction at lower fields when the bias magnetic field is oriented at higher angles. It highlights a complex interplay between the orientation of the magnetic field and the dynamics of magnon modes, likely influenced by factors such as magnetic anisotropy, SW dispersion, demagnetizing fields, and the geometric arrangement of magnetic moments within the material. At an angle of θ = 0°, the edge mode is evenly distributed along the sides. As θ and the external field increase, the edge mode gradually becomes localized on specific sides, creating space for the center mode. When θ is small, a larger external field is needed to localize the edge mode, resulting in a higher coupling field. In contrast, at larger θ values, localizing the edge mode becomes easier, leading to a relatively smaller coupling field, which decreases monotonically as θ increases. At θ = 90°, the center mode becomes too weak or even absent to exchange energy with the dominant edge mode, resulting in a complete absence of coupling. The understanding this critical field and its implications provides valuable insights into the underlying physics of magnonic systems and could guide the design of magnonic devices that exploit this coupling for practical applications.

### 4.2. Analytical Model for Magnon-magnon Coupling

For determination of the coupling strength from the anti-crossing effects as well as to understand the general features of coupling observed in various hybrid quantum systems, several models have been proposed, including coupled harmonic oscillators based on classical physics [22], the dynamic phase correlation model grounded in electromagnetic theory [23], and the quantum mechanical microscopic model [24]. In this work, we adopt the microscopic model based on quantum mechanical theory, as it not only quantitatively describes the observed modes in coupled systems but also can be readily generalized for multi-mode coupling scenarios.

In EMND structure, edge modes are localized near the edges with unique mode profile due to demagnetization effects, while center modes are more uniform in the middle, influenced by the material and internal magnetic field. The internal magnetic field varies within the magnet due to shape anisotropy and the magnetization configuration, causing the edge and center modes to have different frequencies and wavelengths. When these modes match under certain conditions, they can couple and exchange energy, forming hybrid modes. These modes arise from the collective excitation of a large number of ferrimagnetic spins, denoted by $N_s$, within the EMND resonating structure. Since $N_s$ is significantly greater than 1, the Holstein-Primakoff transformation [25] allows us to treat these ferrimagnetic spin excitations as bosonic in nature.

These excitations can be quantized using bosonic operators $\hat{m}_1^\dagger$ and $\hat{m}_1$ for the central mode at resonant frequency $\omega_1$, and $\hat{m}_2^\dagger$ and $\hat{m}_2$ for the edge mode at resonant frequency $\omega_2$. Therefore, the Hamiltonian of the coupled hybrid system can be expressed as:

$$H_s = H_1 + H_2 + H_I \tag{1}$$

Where, $H_1$ represents the Hamiltonian of the central mode, $H_2$ signifies the Hamiltonian of the edge mode arises in the EMND structure and $H_I$ denotes the interaction term that arises due to the coupling between the central and edge magnonic modes. Now, the Hamiltonian of the spin-spin system can be formulated as

$$H_s = \hbar\omega_1 \hat{m}_1^\dagger \hat{m}_1 + \hbar\omega_2 \hat{m}_2^\dagger \hat{m}_2 + \hbar g (\hat{m}_1^\dagger + \hat{m}_1)(\hat{m}_2^\dagger + \hat{m}_2) \tag{2}$$

Here are $\hat{m}_1^\dagger(\hat{m}_1)$ and $\hat{m}_1^\dagger(\hat{m}_1)$ are the creation operators of the distinct ferromagnetic resonance modes that arise at the center and edges of the EMND resonating structure, which have a resonance frequency $\omega_1$ and $\omega_2$, respectively. where $g$ is the coupling strength between both the magnon modes. The interaction term has time dependence of $e^{-i(\omega_c \pm \omega_c)t}$, since most of the experiments form near the coupling center, so rapidly varying term is neglected by rotating wave approximation [26]. Therefore, effective Hamiltonian can be written as

$$H = \hbar\omega_1 \hat{m}_1^\dagger \hat{m}_1 + \hbar\omega_2 \hat{m}_2^\dagger \hat{m}_2 + \hbar g (\hat{m}_1^\dagger \hat{m}_2 + \hat{m}_1 \hat{m}_2^\dagger) \tag{3}$$

The coupling strength results from Zeeman splitting between both resonance modes under an applied magnetic field and is proportional to the square root of the number of spins in the magnetic sample, as predicted by the Tavis-Cummings model [27]. Therefore, increasing the sample size can enhance the coupling strength. However, larger magnetic samples are not suitable for device miniaturization. Alternatively, the coupling strength can be controlled by adjusting the magnetization direction, which can be achieved by altering the direction of the applied magnetic field. From Eq. (3), the equation of motion is,

$$\dot{\hat{m}}_1 = \frac{-i}{\hbar}[\hat{m}_1, H] = -i\omega_1\hat{m}_1 - ig\hat{m}_2 \tag{4a}$$

$$\dot{\hat{m}}_2 = \frac{-i}{\hbar}[\hat{m}_2, H] = -i\omega_2\hat{m}_2 - ig\hat{m}_1 \tag{4b}$$

The above equation describes the intrinsic behavior of a closed system. However, from an experimental perspective, it is essential to account for intrinsic damping, which arises primarily from magnon decay into phonons, spin-wave relaxation, and photon losses. So, intrinsic damping can be included by replacing $\tilde{\omega}_1 = \omega_1 - i\alpha$ and $\tilde{\omega}_2 = \omega_2 - i\alpha$ so the above equations become,

$$\dot{\hat{m}}_1 = -i\tilde{\omega}_1\hat{m}_1 - ig\hat{m}_2 \tag{5a}$$

$$\dot{\hat{m}}_2 = -i\tilde{\omega}_2\hat{m}_2 - ig\hat{m}_1 \tag{5b}$$

Here $\alpha$ is the magnon damping rate. However, Eq. 5 can be solved by considering the initial condition $\hat{m}_1, \hat{m}_2 \propto e^{-i\omega t}$ and solving for the eigenvalues of $\omega$, so the complex eigenfrequencies i.e. $\tilde{\omega}_\pm$ is given by,

$$\tilde{\omega}_\pm = \frac{1}{2}\left[\tilde{\omega}_1 + \tilde{\omega}_2 \pm \sqrt{(\tilde{\omega}_1 - \tilde{\omega}_2)^2 + 4g^2}\right] \tag{6}$$

The frequency of the magnon mode is controlled by the bias magnetic field. As the magnetic field increases, spin waves with different resonance frequencies are generated, and their frequencies shift continuously. When tuned to a specific resonance frequency, $\omega_1=\omega_2$, the modes couple. At this point, the dispersion simplifies to $\omega_\pm = \text{Re}(\tilde{\omega}_\pm) = \omega_1 \pm g$, where small damping is neglected. This coupling results in spin-spin interactions that create a low-energy state with higher stability. Here, $\tilde{\omega}_+$ represents the high-energy state, and $\tilde{\omega}_-$ corresponds to the low-energy state. In order to determine the coupling strength $g$, we fit the Re ($\tilde{\omega}_\pm$) of Eq. (6) to the $|S_{21}|$ power on the plane of f–H as shown in Fig. 4(a), resulting in the emergence of distinct lower- and higher-frequency branches, indicated by the solid black lines. This fitting yielded a *g* value for our EMND resonating structure, ranging from 250 to 380 MHz (Fig. 4(b)), which is comparable to those reported for other nanomagnetic systems in the literature [11, 12]. However, our hybrid system exhibits a higher *g* value than previously studied magnon-magnon coupling in heterostructure or multilayer systems [9,10]. Moreover, the *g* value of our EMND system exceeds that of other hybrid systems, including those involving photon-magnon and phonon-magnon couplings. This highlights the improved coupling capabilities of a single EMND system, which arise from the significant spatial overlap of localized spin-wave modes excited within the same structure.

### 4.3. Control of Coupling Strength

For practical applications of MMC, there is a keen interest in developing methods to control the magnon-magnon interactions. While some studies have explored manipulating this coupling strength by adjusting the geometrical parameters of nanomagnets [11,12] or varying the direction of the applied magnetic field [15,16], investigations into the influence of magnetic properties on magnon-magnon interaction are seldom reported. Gaining a deeper understanding of MMC, as well as the manipulation of coupling strength, could reveal novel phenomena and lead to exciting new applications. Therefore, our next step involves manipulating the magnon-magnon interactions by altering the magnetic properties of EMND system, specifically focusing on intrinsic properties of the magnetic material.

We carried out simulations by varying the damping of EMND from $\alpha = 7.5 \times 10^{-3}$ to $5 \times 10^{-4}$, while keeping all other parameters constant. The resulting $|S_{21}|$ power across the f-H plane for different damping values is illustrated in Fig. 5 (a). It was observed that the value of *g* remains unaffected by changes in damping; however, the intensity of the coupled modes increases, resulting in sharper resonance peaks with a narrower full width at half maximum

Consistent with our findings, the negligible effect of magnetic damping on the coupling strength—defined as the frequency gap between the upper and lower branches of the hybridized modes—has also been reported for other magnetic nanostructures [10-16]. The range of damping parameters used in our simulation aligns with values observed in various materials, such as MgAl ferrite [28], lithium ferrite [29], and CoFeB [30], which exhibit a minimum α value around $10^{-3}$, while YIG [31] can show values as low as $10^{-5}$. Therefore, it is feasible to experimentally observe the strong magnon-magnon coupling confined within a single, finite magnonic cavity.

Alternative approaches to controlling coupling strength include adjusting intrinsic properties of magnetic materials, such as their magnetic anisotropy and saturation magnetization. We conducted simulations for three distinct values of magnetic anisotropy ($K$ = $10^3$, $5\times10^3$ and $10^4$ J/m³). The results (Fig. 5 (b)) indicate that varying $K$ does not affect the strength of magnon-magnon coupling (MMC); however, the coupling center shifts to higher magnetic fields as the magnetic anisotropy field decreases. This shift can be attributed to the dependence of the coupling center on the relative alignment of precession frequencies of each magnon mode, which, in turn, rely on both the external magnetic field and the internal anisotropy field. With a lower anisotropy field, the internal contribution to the effective field diminishes, leading to a decrease in the internal precession frequency. Consequently, a higher external magnetic field is necessary to align the frequencies, reach the resonance condition of the coupled modes, and achieve optimal coupling.

Similarly, the coupling center shifts as a function of saturation magnetization ($Ms$) due to the effect of $Ms$ on the precession frequency of the magnons, which determines the resonance conditions of the system. We performed the simulation by changing the Ms from $8\times10^5$ to $2\times10^6$ and the yielded results are shown in Fig. 5 (c). When $Ms$ increases, the internal precession frequency also increases due to the increase in the internal field. This means that for resonance alignment between coupled modes (to maintain strong coupling), the external magnetic field needed will be lower, effectively shifting the coupling center to lower magnetic fields. Conversely, with lower $Ms$, the internal field decreases, necessitating a higher external magnetic field to reach resonance, and thereby shifting the coupling center to higher fields.

The study reveals that precise tuning of the applied bias field angle, along with careful selection of the intrinsic properties of magnetic nanostructures, substantially influences the magnon-magnon coupling strength of the excited SW modes. This controlled modulation of SW dynamics through bias field adjustments presents an efficient approach for managing SW excitations and MMC strength, paving the way for optimized component design in magnonic circuits and integrated on-chip devices for microwave communication applications. Notably, these results underscore the critical role of material properties in determining MMC strength behavior, where materials with lower magnetic damping demonstrate significantly enhanced SW excitation excitations. This advancement highlights the potential for refined control in magnonic devices, promising substantial improvements in device performance and energy efficiency for advanced communication systems.

## 5. SUMMARY

In summary, we conducted a detailed study on the spin wave (SW) dynamics of elliptical magnonic nano-discs (EMNDs) of $Ni_{80}Fe_{20}$. Our investigation reveals and analyzes magnon-magnon coupling within the EMND system, demonstrating effective control over the anti-crossing phenomenon through the application of an external bias magnetic field. By monitoring microwave excitation power amplitudes, we achieved strong coupling strengths of up to 0.52 GHz. Additionally, we explored the evolution of SW dynamics and anti-crossing by manipulating the magnetic material's intrinsic properties, systematically studying how shifts in coupling centers relate to the applied field angle and magnetic properties. These findings provide new insights into the mechanisms of magnon-magnon coupling, opening avenues for research that could enhance our understanding of spin-wave interactions. This work has significant implications for the design of advanced magnonic devices, where control over coupling centers may enable functionalities like tunable frequency filters, magnetic field sensors, and components for spin-wave-based computing.

**References:**

1. V. V. Kruglyak, S. O. Demokritov, and D. Grundler, "Magnonics" J. Phys. D Appl. Phys. 43, 264001 (2010).

2. A. V. Chumak, V. I. Vasyuchka, A. A. Serga, and B. Hillebrands, "Magnon spintronics" Nat. Phys. 11, 453 (2015).

3. M. Harder, C.-M. Hu, Cavity spintronics "An early review of recent progress in the study of magnon–photon level repulsion", in: R.E. Camley, R.L. Stamps (Eds.), Solid State Physics, vol. 69, Academic Press, 2018, pp. 47–121 (Chapter two).

4. B. Bhoi and S-k Kim "Photon-Magnon Coupling: Historical Perspective, Status, and Future Directions." In the book "Recent Advances in Topological Ferroics and their Dynamics" R. L. Stamps, H. Schultheib (Eds.), Solid State Physics, vol. 70, Academic Press, 2019, pp. 1-77 (Chapter one).

5. B. Zare Rameshti, S. Viola Kusminskiy, J.A. Haigh, K. Usami, D. Lachance-Quirion, Y. Nakamura, C.-M. Hu, H.X. Tang, G.E. Bauer, Y.M. Blanter, Cavity magnonics, Phys. Rep. 979 (2022) 1–61

6. Y. Li, W. Zhang, V. Tyberkevych, W.K. Kwok, A. Hoffmann, V. Novosad, "Hybrid magnonics: physics, circuits, and applications for coherent information processing". J. Appl. Phys. 128(13) (2020).

7. X. Zhang. "A review of common materials for hybrid quantum magnonics". Materials Today Electronics 5 (2023) 100044.

8. S. Klingler et al "Spin-torque excitation of perpendicular standing spin waves in coupled YIG/Co heterostructures" Phys. Rev. Lett. 120 (2018) 127201

9. J. Chen, C. Liu, T. Liu, Y. Xiao, K. Xia, G. E. W. Bauer, M. Wu and H. Yu "Strong interlayer magnon-magnon coupling in magnetic metal-insulator hybrid nanostructures" Phys. Rev. Lett. 120 (2018) 217202

10. Y. Li, et al "Coherent spin pumping in a strongly coupled magnon-magnon hybrid system" Phys. Rev. Lett. 124 (2020) 117202.

11. C. Dai, K. Xie, Z. Pan, and F. Ma "Strong coupling between magnons confined in a single magnonic cavity" J. Appl. Phys. 127 (2020) 203902.

12. Y. Ye, L. Pan, S. Mi, J. Wang, J. Wei, J. Wang and Q. Liu "Strong magnon-magnon coupling in hexagonal magnetic elements" J. Phys. D: Appl. Phys. 57 (2024) 395001.

13. P. K. Pal, A. Barman "Control of magnon-magnon coupling in $Ni_{80}Fe_{20}$ nanocross arrays through system dimensions" J. Magn. Magn. Mater. 588 (2023) 171431.

14. K. Adhikari, S. Sahoo, A.K. Mondal, Y. Otani, and A. Barman, "Large nonlinear ferromagnetic resonance shift and strong magnon-magnon coupling in $Ni_{80}Fe_{20}$" Phys. Rev. B. 101 (2020), 054406.

15. P.K. Pal, S. Majumder, Y. Otani, and A. Barman, "Bias-Field Tunable Magnon-Magnon Coupling in $Ni_{80}Fe_{20}$ Nanocross Array" Adv. Quantum Technol. 6 (2023) 2300003.

16. K. Adhikari, S. Choudhury, S. Barman, Y. Otani, and A. Barman, "Observation of magnon–magnon coupling with high cooperativity in $Ni_{80}Fe_{20}$ cross-shaped nanoring array", Nanotechnology. 32 (2021), 395706.

17. A. Vansteenkiste, J. Leliaert, M. Dvornik, M. Helsen, F. Garcia-Sanchez, and B. Van Waeyenberge, "The design and verification of MuMax3" AIP Adv. 4, (2014) 107133.

18. K.-S. Lee, D.-S. Han, and S.-K. Kim, "Physical Origin and Generic Control of Magnonic Band Gaps of Dipole-Exchange Spin Waves in Width-Modulated Nanostrip Waveguides" Phys. Rev. Lett. 102, 127202 (2009).

19. A. Ghosh, F. Ma, J. Lourembam, J. Xiangjun, R. Maddu, Q. J. Yap, and S. Ter Lim, "Emergent Dynamics of Artificial Spin-Ice Lattice Based on an Ultrathin Ferromagnet" Nano Lett. 20, 109 (2020)

20. P S Keatley, V V Kruglyak, A Neudert, E A Galaktionov, R J Hicken, J R Childress and J A Katine, "Time-resolved investigation of magnetization dynamics of arrays of nonellipsoidal nanomagnets with nonuniform ground states" Phys. Rev. B 78 (2008) 214412

21. L Bai, Y S Gui, Z H Chen, S C Shen, J. Nitta, C-M Hu, L E Hayward, M P Wismayer and B W Southern "Spin wave hybridization via direct mapping of spin wave evolution in ferromagnetic microstructures" J. Appl. Phys. 109 (2011) 093902

22. S. Verma, A. Maurya, R. Singh, B. Bhoi, "Control of Photon-Magnon Coupling in a Planar Hybrid Configuration" J. Supercond. Nov. Magn. 37 (2024)1163–1171.

23. B. Bhoi, B. Kim, S-H Jang, J. Kim, J. Yang, Y-J Cho, and S-K Kim, "Abnormal anticrossing effect in photon-magnon coupling" Phys. Rev. B. 99 (2019) 134426.

24. K. Shrivastava, A. Sahu, B. Bhoi, R. Singh, "Unveiling Electromagnetic Transparency and Absorption in Photon-Photon Coupling" J. Phys. D: Appl. Phys. 57(46) (2024) 465305.

25. T. Holstein, H. Primakoff, Field Dependence of the Intrinsic Domain Magnetization of a Ferromagnet, Phys. Rev. 58 (1940) 1098–1113,

26. C. Gerry, P. Knight, Introductory Quantum Optics, Cambridge University Press, Cambridge, 2004.

27. Tavis, M. & Cummings, F. W. "Exact solution for an N-molecule radiation-field Hamiltonian". Phys. Rev. 170 (1968) 379.

28. S. Emori, et al., "Ultralow Damping in Nanometer-Thick Epitaxial Spinel Ferrite Thin Films" Nano Lett. 18 (2018) 4273–4278.

29. X. Y. Zheng, et al., "Ultra-low magnetic damping in epitaxial Li0.5Fe2.5O4 thin films" Appl. Phys. Lett. 117(2020) 092407.

30. J. Wang et al., "Magnetostriction, Soft Magnetism, and Microwave Properties in Co−Fe−C Alloy Films" Phys. Rev. Appl. 12, 034011 (2019)

31. G. Schmidt, et al., "Ultra-Thin Films of Yttrium Iron Garnet with Very Low Damping: A Review", Phys. Status Solidi B 257 (2020) 1900644.

**Acknowledgments**

The work was supported by the Council of Science & Technology, Uttar Pradesh (CSTUP), (Project Id: 2470, CST, U.P. sanction No: CST/D-1520 and Project Id: 2470, CST, U.P. sanction No: CST/D-1520). B. Bhoi acknowledges support by the Science and Engineering Research Board (SERB) India-SRG/2023/001355. A. Maurya acknowledges the financial support and computational facilities provided by IIT (BHU) Varanasi. S. Verma acknowledges the Ministry of Education, Government of India, for the Prime Minister's Research Fellowship (PMRF ID-1102628).

**Statements & Declarations**

**Funding:** The authors declare that grants were received from CSTUP and SERB during the preparation of this manuscript.

**Competing Interests:** The authors declare that they have no competing interests.

**Author Contributions:** All authors contributed to the study's conception and design. B. B. and R.S. led the work and wrote the manuscript with S.S. K. The other co-authors read, commented, and approved the final manuscript.

**Data Availability:** The data that support the findings of this study are available within the article.

**FIGURE CAPTIONS**

**Fig. 1.** Schematic illustration: simulation set up for elliptical magnonic nano-disc (EMND) and magnon excitation under a bias field $H_{ext}$ and a perturbation field $h_{rf}$.

**Fig. 2** (a) Frequency-field (*f-H*) dispersion spectra of magnons in EMND, presented as color perspective view plots at various out-of-plane angles ($\theta$) of the external magnetic field. (b) Variation of coupling center as a function of $\theta$.

**Fig. 3.** (a) The magnon resonance spectrum and (b) Profies of the FFT power of magnetization components of the two resonance modes for indicated out-of-plane bias field strengths applied at $\theta = 60$.

**Fig. 4.** (a) Planar view of *f-H* dispersion spectra of magnons at various out-of-plane angles ($\theta$) of the external magnetic field. The black solid lines are the results of fts to Eq. 6. (b) Variation of coupling strength 'g' and percentage of change in 'g' as a function of angle $\theta$.

**Fig. 5.** *f-H* dispersion spectra of magnon-magnon coupled modes for various magnetic parameters (a) damping constant, (b) magnetic anisotropy and (c) Saturation magnetization.

**Fig. 1.**

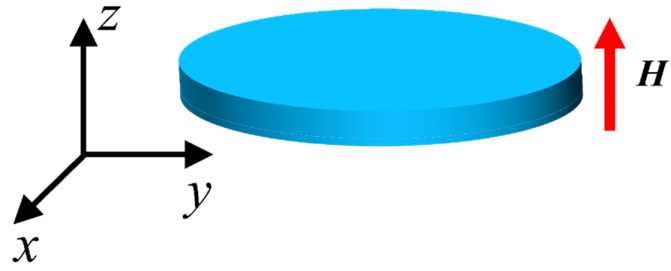

**Fig. 2.**

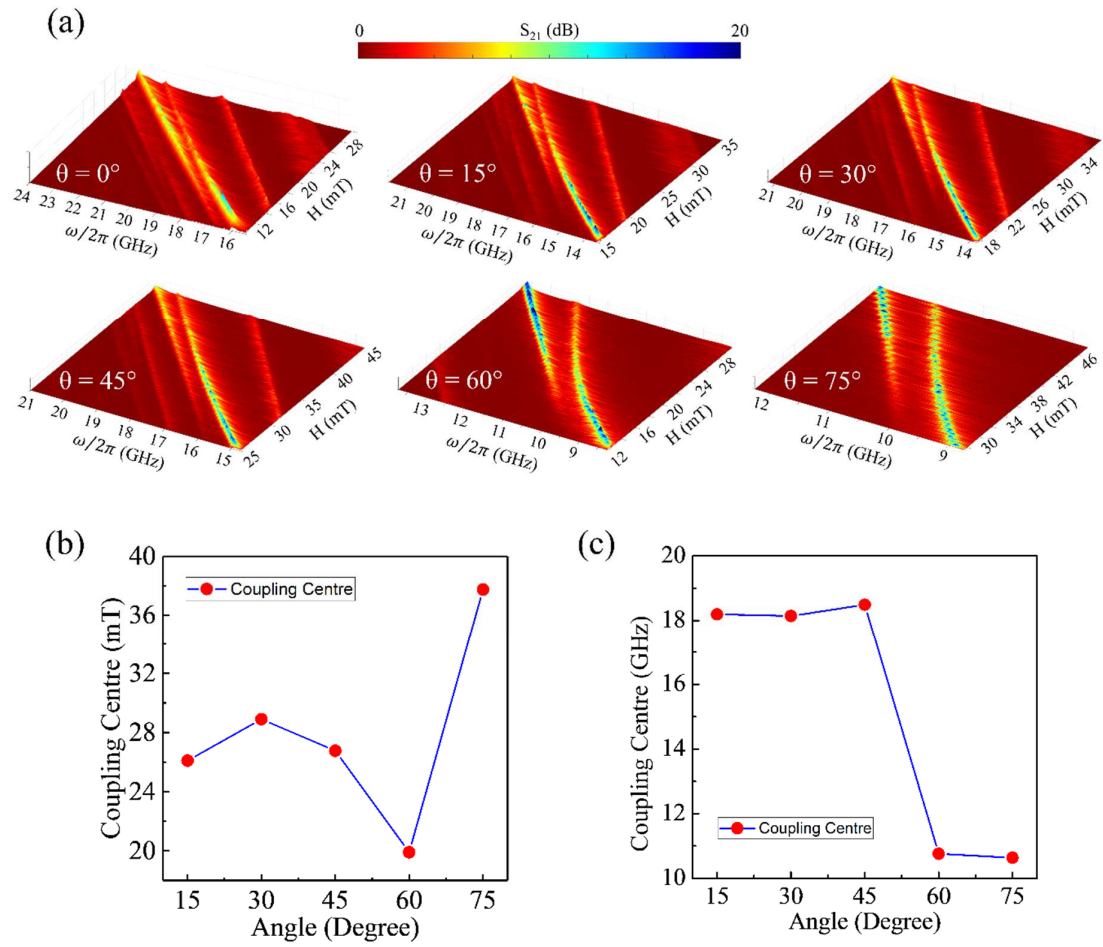

**Fig. 3.**

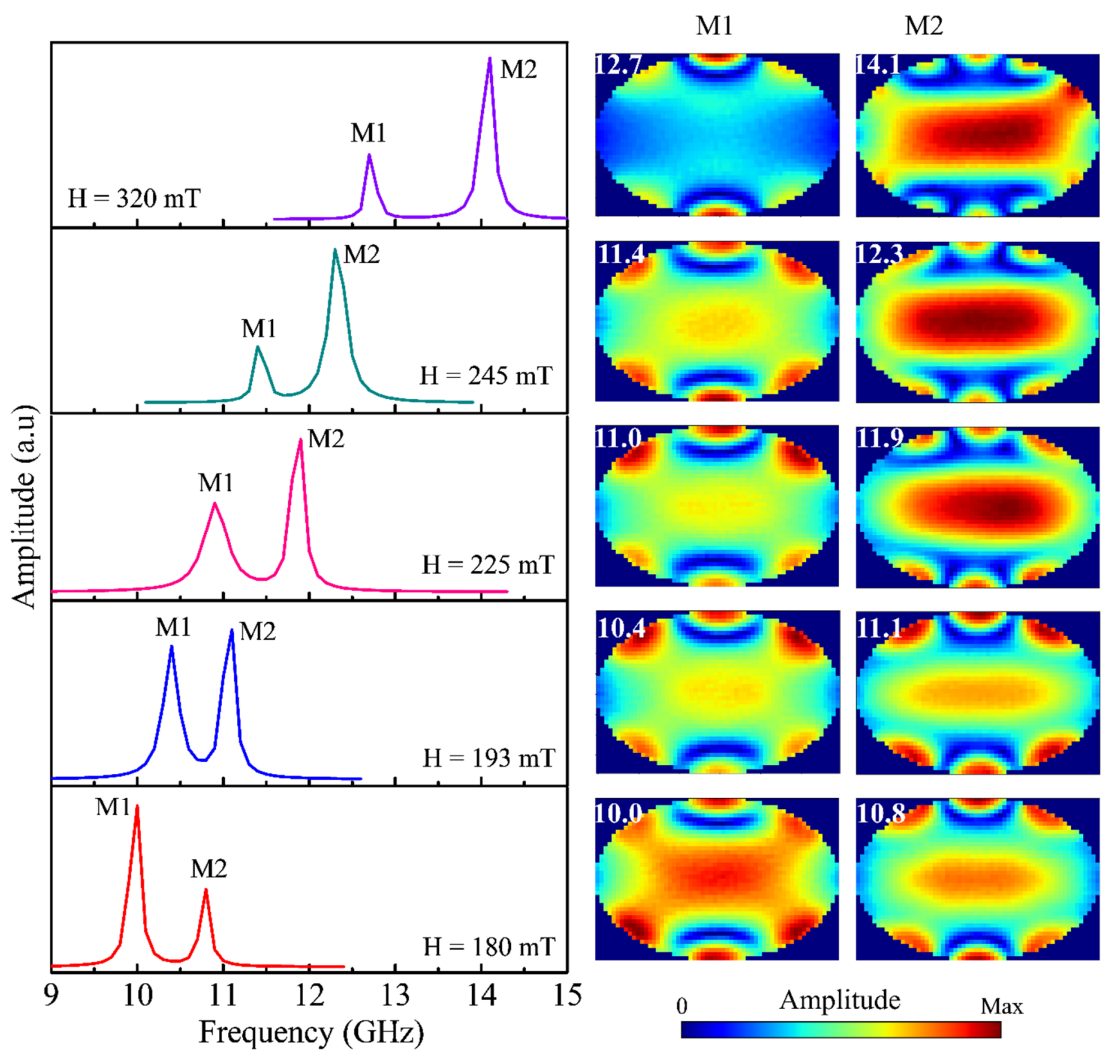

**Fig. 4.**

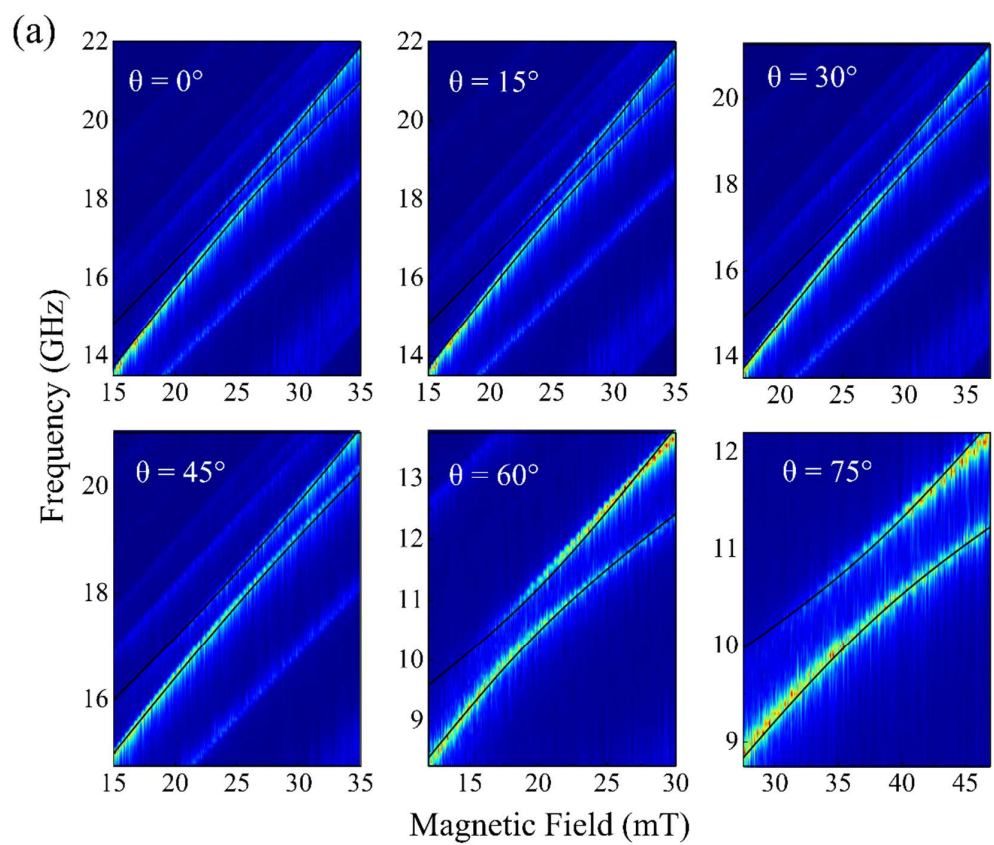

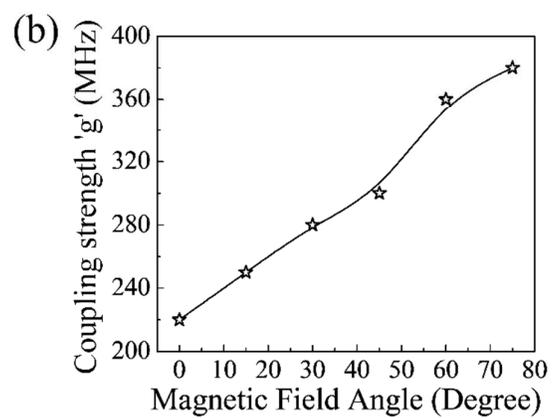

**Fig. 5.**

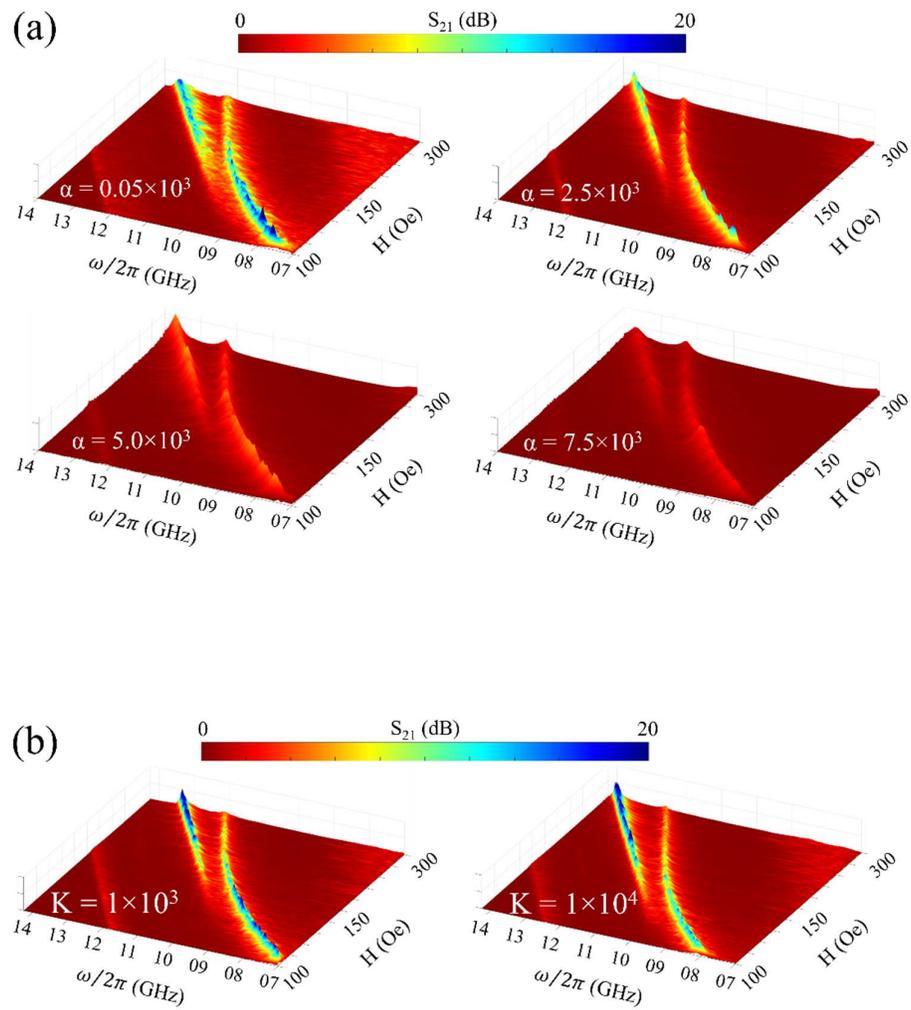